\documentclass[pdftex,twocolumn,epjc3]{svjour3} % twocolumn

\RequirePackage[T1]{fontenc}
\RequirePackage[utf8]{inputenc}
\RequirePackage{graphicx}
\RequirePackage{mathptmx}      % use Times fonts if available on your TeX system
\RequirePackage{flushend}
\RequirePackage[numbers,sort&compress]{natbib}
\RequirePackage[colorlinks,citecolor=blue,urlcolor=blue,linkcolor=blue]{hyperref}
\RequirePackage{amsmath,amssymb} % 常用数学包
\usepackage{stfloats}
\usepackage{booktabs}
\usepackage[protrusion=false]{microtype}

\titlerunning{U-Net-based surrogate modeling for attosecond XFELs}
\authorrunning{Wei et al.}

\begin{document}

\title{U-Net-based surrogate modeling for attosecond X-ray free-electron lasers}

\newcommand{\inst}[1]{\unskip\textsuperscript{#1}}

\author{
  Yufei Wei\inst{1,2}
  \and Bingyang Yan\inst{1,2}
  \and Chenzhi Xu\inst{1,2,5}
  \and Jiawei Yan\inst{4}\textsuperscript{\ensuremath{\dagger}}
  \and Haixiao Deng\inst{3}\textsuperscript{\ensuremath{\ddagger}}
}

\institute{
  \inst{1} Shanghai Institute of Applied Physics, Chinese Academy of Sciences, Shanghai 201800, China.
  \and \inst{2} University of Chinese Academy of Sciences, Beijing 100049, China.
  \and \inst{3} Shanghai Advanced Research Institute, Chinese Academy of Sciences, Shanghai 201210, China.
  \and \inst{4} Deutsches Elektronen-Synchrotron DESY, 22603 Hamburg, Germany
  \and \inst{5} European XFEL, Schenefeld 22869, Germany.
}

\maketitle

\begingroup
\renewcommand{\thefootnote}{\ensuremath{\dagger}}
\footnotetext{Corresponding author: \href{mailto:jiawei.yan@desy.de}{jiawei.yan@desy.de}}
\renewcommand{\thefootnote}{\ensuremath{\ddagger}}
\footnotetext{Corresponding author: \href{mailto:denghx@sari.ac.cn}{denghx@sari.ac.cn}}
\endgroup

% \date{Received: date / Accepted: date}
% The correct dates will be entered by the editor

\begin{abstract}

Attosecond X-ray pulse generation in modern X-ray free-electron lasers relies on strongly compressed, precisely tailored electron bunches, making accurate diagnostics and control of the two-dimensional longitudinal phase space (LPS) essential. In the self-chirping scheme, collective effects in the linac generate a strong energy chirp that is converted into high peak current through pre-undulator compression, enabling isolated attosecond pulse generation. Reliable operation of this scheme depends on precise two-dimensional LPS control and fast diagnostics. In this work, we present a U-Net-based neural network surrogate that predicts two-dimensional LPS distributions directly from accelerator settings. The model exhibits good agreement with start-to-end simulation results. These results demonstrate the potential of neural-network surrogates to facilitate fast virtual diagnostics and provide a step toward real-time tuning in attosecond XFEL operation.

\end{abstract}

\section{Introduction}

Relativistic electron beam driven light sources, including synchrotron radiation facilities and X-ray free-electron lasers (XFELs), are essential tools across physics, chemistry, materials science, and structural biology. Modern XFELs deliver ultrashort, high brightness X-ray pulses that enable time-resolved measurements on atomic length scales and femtosecond time scales~\cite{huang2021features}. A central frontier is to extend this control to the attosecond regime and to deliver user relevant pulse formats with reliable shot-to-shot stability. Achieving that goal requires electron beam phase space control with increasingly demanding precision.

In recent years, a wide range of attosecond XFEL concepts has been proposed and demonstrated by shaping the electron beam longitudinal phase space (LPS) to create a narrow, high current region that dominates the FEL interaction. For example, the self-chirping mode at the European XFEL demonstrated terawatt-scale hard X-ray pulses \cite{yan2024terawatt}. Moreover, the AttoSHINE project has been proposed to generate high-power attosecond XFEL with continuous-wave XFELs, such as the Shanghai High Repetition Rate XFEL and Extreme Light Facility (SHINE) \cite{Yan2025}.

Realizing and stabilizing such LPS-sensitive XFEL modes requires iterative optimization of machine settings and frequent feedback on the resulting phase space. A conventional approach is to rely on computationally intensive start-to-end simulations to scan the relevant parameter space, whereas such simulations are often time-consuming and resource-demanding. Experimentally, the two-dimensional LPS can be measured using an X-band transverse deflecting cavity (TCAV) combined with a dipole magnet, which streaks and disperses the bunch onto a screen to produce single-shot phase-space images~\cite{yan:fel2019-thp018}. While highly informative, TCAV-based diagnostics are inherently invasive and therefore cannot be deployed continuously during user operation, especially at high repetition rates. These limitations motivate fast surrogate approaches for two-dimensional LPS diagnostics and phase-space design.

Machine learning has become a practical tool for building data-driven surrogates in accelerator physics, ranging from Gaussian process regression to deep neural networks \cite{10.7551/mitpress/3206.001.0001,LeCun2015DeepLearning,PhysRevAccelBeams.23.044601,zhu:fel2022-wep12}. By learning nonlinear mappings directly from archived or actively sampled data, surrogate models provide fast approximations of computationally expensive simulations or experimentally inaccessible diagnostics, and have been successfully used for high-dimensional control problems~\cite{Roussel2024PRAB}. Multi-objective Bayesian optimization has been applied on operating machines to tune beams and map trade-offs between pulse energy, bandwidth, and beam quality~\cite{Roussel2021PRAB,Ji2024NatCommun_MOBO,Xu2025IPAC_MOPB038}. Complementary approaches, including evolutionary many-objective optimization, have also been explored~\cite{Yan2019PRAB}. In particular, neural-network surrogates have demonstrated shot-to-shot prediction of electron-beam two-dimensional LPS with good agreement to deflector-based measurements~\cite{Emma2018,PhysRevApplied.16.024005}, and similar approaches have been applied to infer photon pulse properties from fast diagnostics~\cite{sanchez2017accurate}.

In this work, we develop a U-Net based neural network surrogate that predicts two dimensional LPS distributions directly from accelerator settings in the AttoSHINE. The model is trained on start-to-end simulations and evaluated using both image based and physics motivated metrics. The resulting surrogate provides a fast forward model for parameter scans toward target phase space structures, and it is a step toward real time, two-dimensional LPS aware tuning workflows for high peak power attosecond XFEL operation.

\begin{figure}[htbp]
    \centering
    \includegraphics[width=0.5\textwidth]{unet.png}
    \caption{Schematic of the modified U-Net architecture. Feature maps are annotated as $C\times H\times W$. The encoder uses $2\times2$ max pooling followed by residual blocks. The bottleneck includes an ASPP module. The decoder uses bilinear upsampling and attention-gated skip fusion. Encoder features are concatenated with decoder features at matching scales, and padding is applied only when size alignment is required. A final $1\times1$ convolution projects decoder features to the output density map.}
    \label{unet}
\end{figure}

\section{Methods}

U-Net is a fully convolutional encoder decoder architecture with skip connections, originally proposed by Ronneberger \emph{et al.} for medical image segmentation~\cite{Ronneberger2015UNet}. By combining global context learned in the contracting path with high resolution features delivered through skip connections, U-Net is well suited for dense map prediction tasks that require both global structure and fine detail recovery. In this work, we adopt a modified U-Net style backbone augmented with residual blocks~\cite{He2016DeepResidual}, attention-gated skip fusion~\cite{Oktay2018AttentionUNet}, and multi-scale context aggregation via atrous spatial pyramid pooling (ASPP)~\cite{chen2017deeplabsemanticimagesegmentation}.

The network architecture used in this study is shown in Fig.~\ref{unet}. The three accelerator phases form a length-three conditioning vector, and each phase is broadcast to a constant two-dimensional feature map of size \(256\times256\) so that the input can be processed by fully convolutional layers. We further concatenate explicit coordinate channels and sinusoidal positional encodings. Specifically, the normalized coordinates $(x,y)\in[-1,1]$ are appended as two additional channels, and Fourier-style encodings are constructed using $\sin(\pi f x)$, $\cos(\pi f x)$, $\sin(\pi f y)$, and $\cos(\pi f y)$ at frequencies $f\in\{1,2,4,8\}$~\cite{Liu2018CoordConv,Tancik2020FourierFeatures}. As a result, the final network input has dimension \(21\times256\times256\), including three broadcast phase maps, two coordinate channels, and sixteen positional-encoding channels. The resulting tensor is fed into a symmetric encoder-decoder with four downsampling stages and a bottleneck.

Each encoder stage applies a $2\times2$ max pooling operation followed by a residual block comprising two $3\times3$ convolutions, Group Normalization, and ReLU activations~\cite{Wu2018GroupNorm}, together with an identity skip connection. When the number of channels changes, a $1\times1$ projection is used on the residual branch to match dimensions~\cite{He2016DeepResidual}. At the bottleneck, an ASPP module composed of parallel atrous convolutions with multiple dilation rates expands the effective receptive field and captures multi scale context~\cite{chen2017deeplabsemanticimagesegmentation,Chen2017DeepLabv3}. The decoder mirrors the encoder with four upsampling stages using bilinear upsampling, progressively restoring the resolution back to $256\times256$. At each decoder level, the corresponding encoder feature map is filtered by an attention gate before concatenation, enabling selective use of high resolution cues from the skip pathway~\cite{Oktay2018AttentionUNet}. A final $1\times1$ convolution produces the main prediction, matching the three channel representation used during training, where the same density map is replicated across channels for implementation convenience.

In addition to the main output, we introduce an auxiliary edge-prediction head attached to an intermediate decoder feature map to provide explicit structural supervision during training. The supervision target for this branch is obtained by applying a Sobel-type gradient operator to the reference density map. We also apply deep supervision by attaching auxiliary prediction heads to multiple decoder scales, whose outputs are upsampled to the full resolution and supervised against the target image during training.

The overall training objective is a structure aware loss
\begin{equation}
L
=
L_{\mathrm{rec}}
+
\lambda_g L_{\mathrm{grad}}
+
\lambda_e L_{\mathrm{edge}}
+
\lambda_{ds} L_{\mathrm{ds}},
\end{equation}
where $L_{\mathrm{rec}}$ is a pixel wise reconstruction loss (L1) between prediction and target, $L_{\mathrm{grad}}$ enforces gradient consistency by comparing Sobel gradient magnitude maps of prediction and target, $L_{\mathrm{edge}}$ supervises the auxiliary edge head against the target gradient magnitude map, and $L_{\mathrm{ds}}$ is an average L1 loss over the multi scale auxiliary outputs. In our implementation we set $(\lambda_g,\lambda_e,\lambda_{ds})=(3.0,1.0,0.5)$. The gradient based terms encourage the model to focus on structural discrepancies such as phase space boundaries rather than only matching intensity offsets.

All models were optimized using AdamW and trained on a single NVIDIA A100 GPU, without distributed or multi-GPU training. Training ran for 500 epochs with an initial learning rate of $1 \times 10^{-5}$. If the validation loss failed to improve for 10 consecutive epochs, the learning rate was multiplied by a factor of 0.5, with a lower bound (minimum learning rate) set to $1 \times 10^{-7}$.

\begin{figure*}[!t]
  \centering
  \includegraphics[width=\textwidth]{attosecond_scheme_v2_1.jpg} 
\caption{Schematic layout of the self-chirping process in AttoSHINE and the corresponding evolution of the electron-beam LPS. Insets show representative LPS distributions (colored density) together with the current profile (gray curve) at three locations along the beamline. (a) After nonlinear compression downstream of BC2. (b) At the end of L4. (c) After further compression in LTU1. The arrows labeled a--c indicate the corresponding observation points. The head of the electron beam is on the left.}
  \label{example1}
\end{figure*}

\section{Results of electron-beam prediction}

To demonstrate two-dimensional LPS prediction capability, we consider the electron-beam manipulation in AttoSHINE. The layout and the corresponding two-dimensional LPS evolution during the self-chirping scheme are shown in Fig.~\ref{example1}. The layout consists of the linac sections L1, LH, L2, L3, and L4, the bunch-compression chicanes BC1 and BC2, the beam transport line LTU1, and the undulator section FEL1. As an illustrative example, we adopt a two-stage compression scheme. By tuning the RF phases in L1, LH, and L2, the bunch undergoes nonlinear compression after BC2, as shown in Fig.~\ref{example1}(a). By the end of the linac section L4, a pronounced energy chirp has developed along the bunch, as shown in Fig.~\ref{example1}(b). After further compression in LTU1, the bunch length is significantly shortened, as shown in Fig.~\ref{example1}(c), which is favorable for generating attosecond pulses. In this work, we take the two-dimensional LPS at location (b) as the prediction target and model the tuning of the RF phases in L1, LH, and L2. Throughout the simulations, the momentum compaction of the compression system is kept fixed. Starting from an optimized baseline case, we scan the three RF phases within specified ranges. The scan ranges for each parameter are summarized in Table~\ref{tab:input_ranges}. This parameter scan yields a dataset of 10200 simulations performed with ELEGANT using \(10^{5}\) macroparticles per run. From the resulting samples, 8,000 are used for training, 2,000 for validation, and 200 are reserved as an independent test set. All results below are evaluated on the independent 200-case test set. 

\begin{figure*}[!t]
    \centering
    \includegraphics[width=\textwidth]{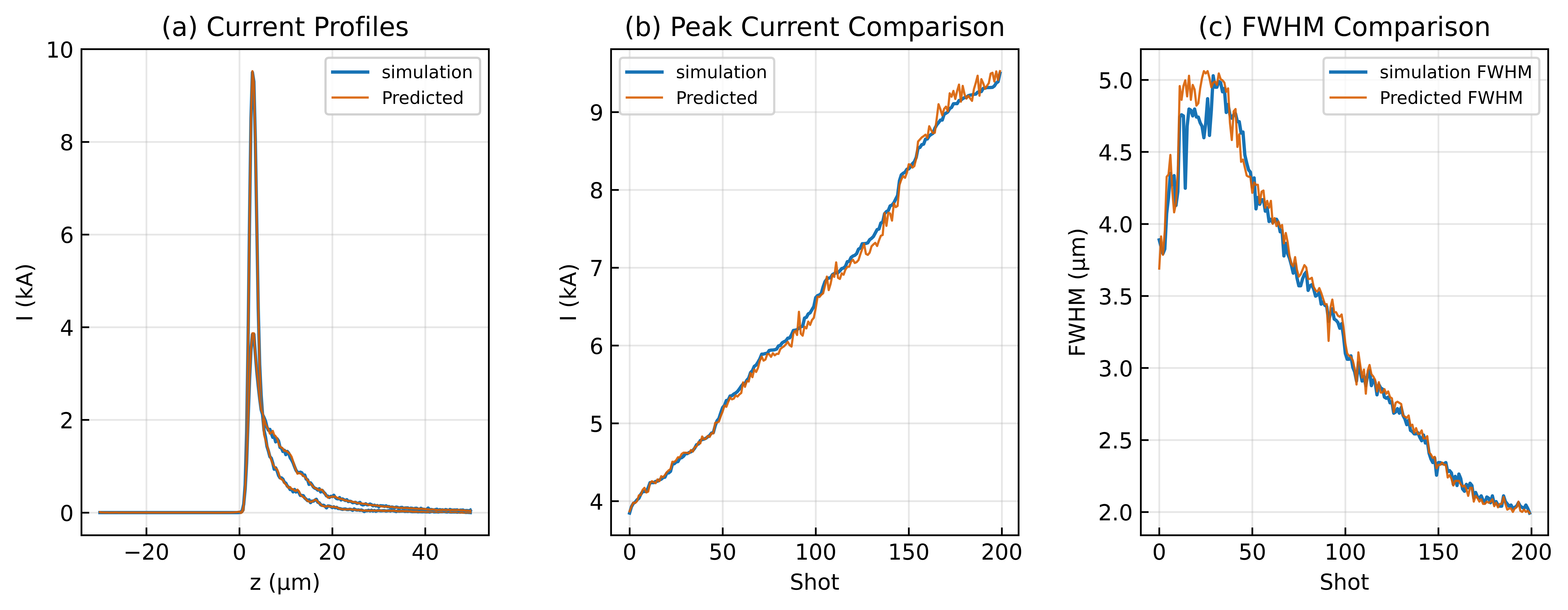}
    \caption{Comparison of simulated and U-Net-predicted current profiles derived from the LPS at the end of L4 (observation point (b) in Fig.~\ref{example1}) for the 200-case test set. (a) Representative reconstructed current profiles (including examples with the highest and lowest peak currents in the selected set). (b) Shot-by-shot comparison of peak current with shots sorted by the simulated peak current. (c) Shot-by-shot comparison of the full width at half maximum for the same shot ordering as in (b), showing the corresponding FWHM values for the peak-current-sorted cases. The close overlap between prediction and simulation indicates good agreement.}
    \label{fig3}
\end{figure*}

\begin{table}[ht]
  \centering
  \begin{tabular}{l c}
    \toprule
    Parameters & Range \\
    \midrule
    L1 phase & [-4.78, -4.38] \\
    LH phase & [-142.61, -141.81] \\
    L2 phase & [-33.64, -33.24] \\
    \bottomrule
  \end{tabular}
  \caption{Scan ranges of the three RF phase settings used as input features in the simulations. The phases of L1, LH, and L2 are varied while other lattice settings are kept fixed.}
  \label{tab:input_ranges}
\end{table}

Figure~\ref{fig3} compares simulated and predicted current profiles obtained by projecting the LPS at the end of L4, onto the longitudinal axis. The neural-network predictions reproduce the simulated current profiles well, as illustrated by the representative example in Fig.~\ref{fig3}(a). Shot-by-shot comparisons of the peak current and the full width at half maximum (FWHM) between simulation and prediction show good agreement, as shown in Fig.~\ref{fig3}(b) and (c). Across all tested cases, the mean absolute relative differences between the predicted and simulated profiles are 0.95\% for the peak current and 1.77\% for the FWHM.

\begin{figure}[htbp]
    \centering
    \includegraphics[width=0.5\textwidth]{nmae_r2_histogram.png}
    \caption{Histogram of the NMAE and $R^2$ for 200 test set cases.}
    \label{nmae}
\end{figure}

To provide a more physically intuitive and quantitatively rigorous validation, we evaluate the reconstruction accuracy using the normalized mean absolute error (NMAE) and the $R^2$ score between the predicted images $\hat{Y}$ and the ground-truth simulated images $Y$. They are defined as
\begin{equation}\label{eq:nmae}
\mathrm{NMAE} \;=\; 100 \times
\frac{\displaystyle\sum_{i,j}\left|Y_{ij}-\hat{Y}_{ij}\right|}
     {\displaystyle\sum_{i,j}\left|Y_{ij}\right|},
\qquad
R^2 \;=\;
1 - \frac{\displaystyle\sum_{i,j}\left(Y_{ij} - \hat{Y}_{ij}\right)^2}
         {\displaystyle\sum_{i,j}\left(Y_{ij} - \bar{Y}\right)^2},
\end{equation}
where $i,j\in\{1,\dots,256\}$ index pixel positions and $\bar{Y}$ denotes the mean of $Y_{ij}$ over all pixels for each simulated image. The resulting per-sample NMAE and $R^2$ statistics for 200 test-set cases are summarized in Fig.~\ref{nmae}. The mean ($\pm$ rms) NMAE is $13.83\% \pm 0.84\%$, and the mean ($\pm$ rms) $R^2$ is $0.9866 \pm 0.0016$.

% To provide a more physically intuitive and quantitatively rigorous validation, we use the normalized mean absolute error (NMAE) between the predicted images $\hat{Y}$ and the ground-truth simulated images $Y$ to quantify reconstruction accuracy. The NMAE is defined as
% \begin{equation}\label{eq:nmae}
% NMAE \;=\; 100 \times
% \frac{\displaystyle\sum_{i}\sum_{j}\left|Y_{ij}-\hat{Y}_{ij}\right|}
%      {\displaystyle\sum_{i}\sum_{j}\left|Y_{ij}\right|},
% \end{equation}
% where $i,j\in\{1,\dots,256\}$ index pixel positions. The resulting per-sample NMAE statistics for 200 test set cases are summarized in Fig.~\ref{nmae}. The mean NMAE is 13.83\%.

\begin{figure*}[!t]
    \centering
    \includegraphics[width=\textwidth]{three_nmae_sorted_shots_composite.png}
    \caption{Three representative test-set examples. For each case, the left panel shows the simulated longitudinal phase space, the middle panel shows the U-Net prediction, and the right panel compares the corresponding current profiles from simulation and prediction. From top to bottom, the NMAE values in phase space are 12.38\%, 13.44\%, and 18.21\%, respectively.}
    \label{three_exampels}
\end{figure*}

Figure~\ref{three_exampels} presents three representative test-set examples with NMAE values of 12.38\%, 13.44\%, and 18.21\%, corresponding to the lowest-error case, a near-mean case, and the highest-error case, respectively. Reconstructions near the mean error level ($\sim$13\%) reproduce the LPS structure with good visual fidelity. Even in the highest-error example (18.21\%), the prediction preserves the dominant phase-space morphology, including the main high-current feature and fine-scale structures, and the resulting current profile remains qualitatively consistent with the start-to-end simulation, supporting its use for diagnostic interpretation.

In terms of computational efficiency, a single start-to-end ELEGANT simulation requires about \(353\,\mathrm{s}\) on a laptop CPU (Intel(R) Core(TM) i7-10700, single core). By contrast, a single surrogate evaluation takes approximately \(150\,\mathrm{ms}\) on the same CPU core. Although the wall-clock time of start-to-end simulations can be reduced by parallel execution on multi-core clusters, high-fidelity studies often require macroparticle numbers in the \(10^{6}\) range (or higher) to suppress statistical noise, especially when resolving fine structures and low-density regions. This requirement increases the per-run simulation cost and further widens the computational advantage of the surrogate model for parameter scans and iterative tuning workflows.

\section{Ablation experiment}

\begin{figure}[htbp]
    \centering
    \includegraphics[width=0.4\textwidth]{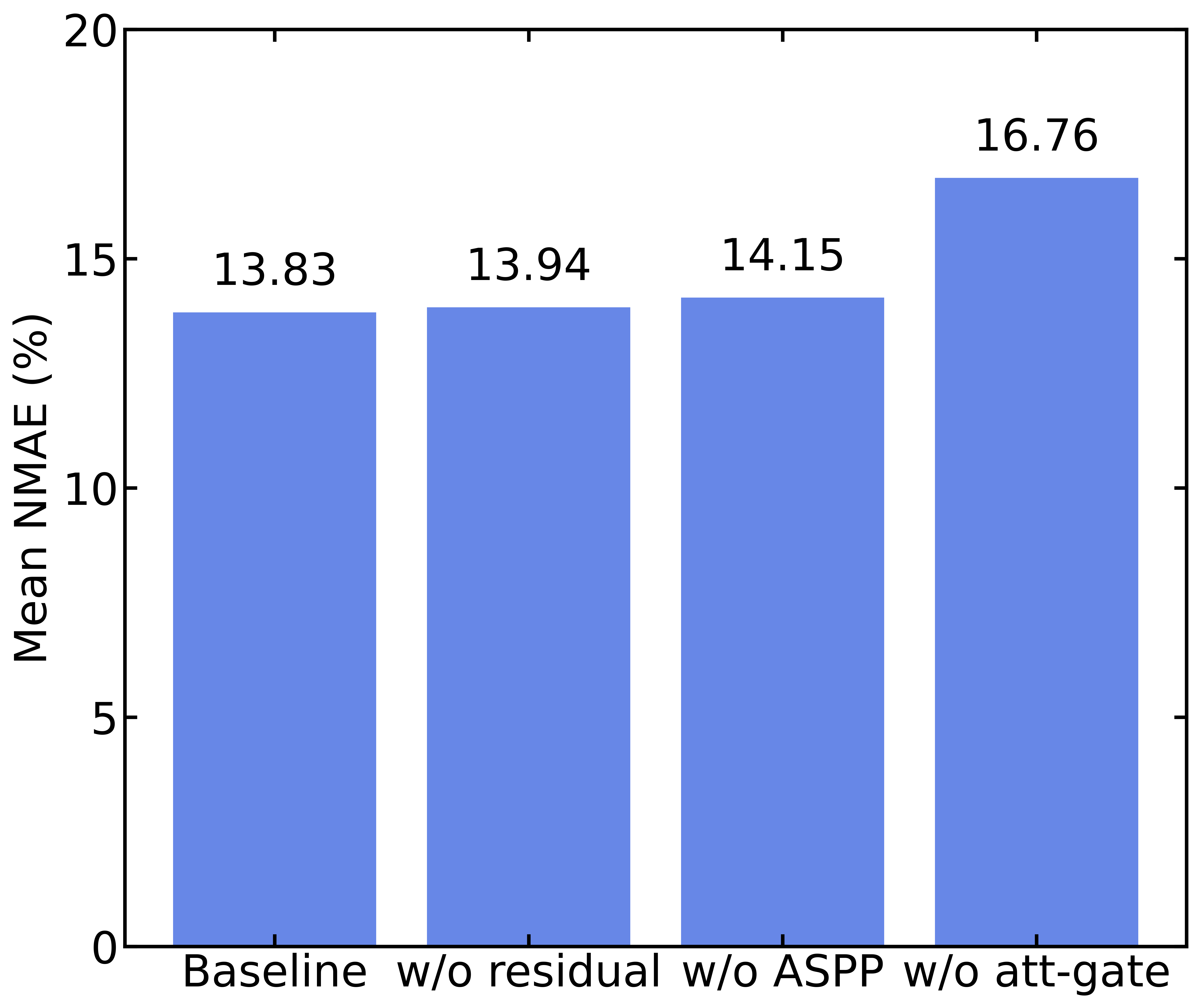}
\caption{Mean NMAE on the test set for the baseline model and the three architectural ablation variants, namely without attention-gated skip fusion, without ASPP, and without residual connections.}
    \label{arch_ablation}
\end{figure}

To evaluate the contribution of different components, a series of controlled ablation experiments was performed. We first examined three architectural modules, including the attention-gated skip fusion, the ASPP module, and the residual connections. In each ablation, only the target component was modified, while all other settings, including the dataset split, optimizer, learning-rate schedule, loss function, and training procedure, were kept identical to those of the baseline model. Specifically, the ASPP ablation removes the ASPP module at the bottleneck, the attention-gated skip-fusion ablation replaces the attention gate with an identity mapping, and the residual ablation replaces residual blocks with plain convolutional blocks while preserving the overall encoder and decoder structure. The corresponding results are shown in Fig.~\ref{arch_ablation}. Among these variants, removing the attention-gated skip fusion causes the largest degradation, with the mean NMAE increasing from 13.83\% for the baseline model to 16.76\%. By contrast, removing ASPP and residual connections leads to much smaller changes, yielding mean NMAE values of 14.15\% and 13.94\%, respectively. These results indicate that, under the present setting, attention-gated skip fusion provides the largest performance gain among the architectural refinements examined here.

\begin{figure}[htbp]
    \centering
    \includegraphics[width=0.4\textwidth]{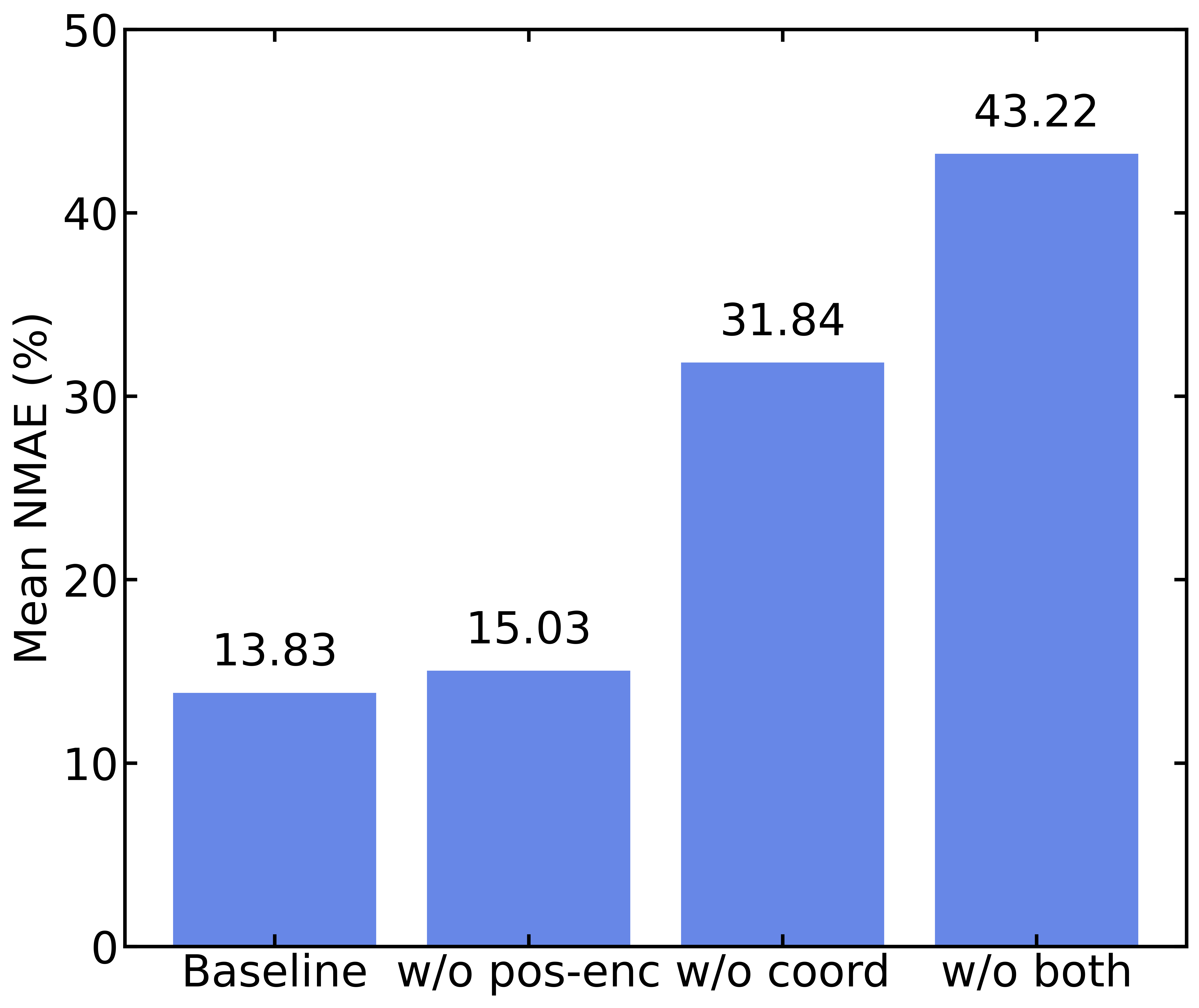}
    \caption{Mean NMAE on the test set for the baseline model and the positional-representation ablation variants, including without sinusoidal positional encodings, without coordinate channels, and without both components.}
    \label{pos_ablation}
\end{figure}

The role of positional representation was then examined. This representation consists of explicit coordinate channels and sinusoidal positional encodings. Three variants were considered, namely removing the positional encodings only, removing the coordinate channels only, and removing both simultaneously. As shown in Fig.~\ref{pos_ablation}, removing the positional encodings alone increases the mean NMAE from 13.83\% to 15.03\%. Removing the coordinate channels results in a much larger degradation, with the mean NMAE increasing to 31.84\%. When both components are removed, the mean NMAE further rises to 43.22\%, which is the worst result among all tested variants. These results show that positional information is essential for this task, and that the coordinate channels provide the dominant contribution to the reconstruction performance. Figure~\ref{pos_ablation_shot} shows an example for the model without both coordinate channels and positional encodings. Compared with the simulated phase-space distribution, the prediction exhibits substantial distortion and unphysical negative regions. The corresponding current profile also shows a clear deviation from the reference. This behavior further indicates that explicit positional information is important for preserving spatial consistency in the reconstructed phase space.

\begin{figure*}[!t]
    \centering
    \includegraphics[width=\textwidth]{pos_ablation_worst_reconstruction_pub.png}
    \caption{An example for the model without both coordinate channels and positional encodings, with an NMAE of 64.83\%. The left panel shows the simulated longitudinal phase space, the middle panel shows the predicted phase space, and the right panel compares the corresponding current profiles from simulation and prediction.}
    \label{pos_ablation_shot}
\end{figure*}

\section{Conclusions}

We have presented a U-Net-based neural-network surrogate for predicting two-dimensional electron-beam LPS distributions from a small set of accelerator phase settings in the AttoSHINE self-chirping scheme. Trained on start-to-end simulation data, the model reproduces the main LPS morphology with good agreement and accurately captures the projected current profiles relevant to attosecond XFEL operation. In particular, the surrogate achieves low inference latency while preserving the key phase-space structures associated with high-current spike formation, indicating its potential as a fast forward model for virtual diagnostics and rapid parameter scans. The ablation studies further clarify the role of different architectural components. Among the tested architectural refinements, attention-gated skip fusion provides the largest improvement in reconstruction accuracy, whereas explicit positional information, especially the coordinate channels, plays a more dominant role in preserving the spatial consistency of the predicted phase-space distributions.

At the same time, it should be emphasized that the present results are obtained within a relatively narrow parameter range around a fixed operating point, with only three RF phase settings varied while other machine parameters are kept unchanged. The current model should therefore be regarded as a local surrogate for this specific attosecond operation scenario, rather than a fully general model covering the full accelerator parameter space. Future work will extend the input space to broader operating regimes and additional control variables, such as more RF phases, accelerating gradients, and other relevant settings, in order to improve generalization and better support realistic online tuning applications.

Further developments will also require experimental validation, as well as bridging the gap between simulation and realistic accelerator measurements. In practice, prediction performance may be affected by distribution shifts, machine drifts, and unmodeled variations between simulation and real operation. Addressing these issues will require uncertainty quantification and improved robustness across different machine conditions. With these advances, AI-based surrogate models could become an enabling layer for efficient commissioning, stabilization, and optimization of LPS-sensitive attosecond XFEL modes at high repetition rates.

\section{Acknowledgements}

This work was supported by the National Natural Science Foundation of China (12125508, 12541503), the National Key Research and Development Program of China~(2024YFA1612104) and Shanghai Pilot Program for Basic Research – Chinese Academy of Sciences, Shanghai Branch (JCYJ-SHFY-2021-010).  Jiawei Yan acknowledges support from DESY (Hamburg, Germany), a member of the Helmholtz Association (HGF), and the European XFEL (Schenefeld, Germany).

\section{Data Availability Statement }
Data supporting the findings of this study are available from the corresponding author upon reasonable request.

\bibliographystyle{unsrt}
\bibliography{refs}

\end{document}